\newcommand{\be}{\begin{equation}}
\newcommand{\ee}{\end{equation}}
\newcommand{\bea}{\begin{eqnarray}}
\newcommand{\eea}{\end{eqnarray}}
\newcommand{\nn}{\nonumber\\}
\newcommand{\p}[1]{(\ref{#1})}
\newcommand{\lb}{\label}

\newcommand{\cD}{{\cal D}}
\newcommand{\cDb}{{\bar{\cal D}}}
\newcommand{\cW}{{\cal W}}
\newcommand{\cWb}{{\bar{\cal W}}}
\documentstyle[12pt]{article}

\begin{document}
\begin{titlepage}
\begin{flushright}
LNF-00/040(P) \\
JINR-E2-2000-311 \\
hep-th/0012236\\
December 2000
\end{flushright}
\vskip 0.6truecm
\begin{center}
{\Large\bf
N=2 and N=4 supersymmetric Born-Infeld} \\

\vspace{0.2cm}
{\Large\bf theories from nonlinear realizations}
\end{center}
 \vskip 0.6truecm
\centerline{ S. Bellucci${}^{\,a,1}$, E. Ivanov${}^{\,b,2}$,
S. Krivonos${}^{\,b,3}$ }

\vskip 0.6truecm
\centerline{${}^a${\it INFN-Laboratori Nazionali di Frascati,}}
\centerline{\it P.O.Box 13, I-00044 Frascati, Italy}

\vspace{0.5cm}
\centerline{${}^b${\it Bogoliubov Laboratory of
Theoretical Physics, JINR,}}
\centerline{\it 141 980 Dubna, Moscow region,
Russian Federation}

\vspace{0.2cm}
\vskip 0.6truecm  \nopagebreak

\begin{abstract}
\noindent Starting from nonlinear realizations of 
the partially broken central-charge extended $N=4$ and 
$N=8$ Poincar\'e supersymmetries in $D=4$, 
we derive the superfield equations of $N=2$ and $N=4$
Born-Infeld theories. The basic objects are the 
bosonic Goldstone $N=2$ and $N=4$ superfields associated 
with the central charge generators.
By construction, the equations are manifestly $N=2$ and $N=4$ 
supersymmetric and enjoy covariance under
another nonlinearly realized half of the original supersymmetries. 
They provide a manifestly worldvolume
supersymmetric static-gauge description of D3-branes in $D=6$ and $D=10$.
For the $N=2$ case we find, to lowest orders, the equivalence
transformation to the standard $N=2$ Maxwell superfield strength 
and restore, up to the sixth order, the off-shell $N=2$ Born-Infeld 
action with the second hidden $N=2$ supersymmetry.
\end{abstract}

\vfill

\noindent{\it E-Mail:}\\
{\it 1) bellucci@lnf.infn.it}\\
{\it 2) eivanov@thsun1.jinr.ru}\\
{\it 3) krivonos@thsun1.jinr.ru}
\newpage

\end{titlepage}

\renewcommand{\thefootnote}{\arabic{footnote}}
\setcounter{footnote}0
\setcounter{equation}0
\section{Introduction}
Supersymmetric extensions of the Born-Infeld (BI)
theory in diverse dimensions are currently under intensive study.
This vast interest is mainly due to the fact that the
corresponding actions naturally arise in string theory as
the worldvolume actions of Dp-branes (see \cite{Tseyt}
and refs. therein). It is of importance to have superfield
formulations of supersymmetric BI theories, with all linearly realized
supersymmetries being manifest.

The $N=1, D=4$ super BI action constructed in \cite{CF}
is not entirely determined by $N=1$ supersymmetry.
As was shown in \cite{BG2}, this freedom can be fixed by
requiring the action to possess an extra
hidden nonlinearly realized $N=1$ supersymmetry completing
the manifest one to $N=2, D=4$ supersymmetry. Such a form of the $N=1$
BI action naturally comes out within the nonlinear realization of
$N=2$ supersymmetry partially broken to $N=1$, with the vector gauge
$N=1$ multiplet as the Goldstone one. In components,
the action is reduced to the static-gauge form of the worldvolume
action of the ``space-filling'' D3-brane \cite{BG2,RT}.

Until now, the action of refs. \cite{BG2,RT} remains the only example of
a superfield BI action in $D\geq 4$ having
both manifest and hidden supersymmetries, and so admitting
an interpretation as the Goldstone superfield
action (or a worldvolume-supersymmetric form
of some D3-brane action). The $N=2$ BI action
constructed in \cite{Ket} possesses no hidden supersymmetry \cite{SKT}.
So it can be viewed merely as a part of the as yet unknown action of
the Goldstone $N=2$ vector multiplet supporting the spontaneous
partial breaking $N=4 \rightarrow N=2$ as
was suggested in \cite{BIK1,BIK2}. As for $N=4$ BI action
which is a candidate for the Goldstone multiplet action associated with  
the breaking $N=8 \rightarrow N=4$ in $D=4$
(the static-gauge action of D3-brane in $D=10$ \cite{Ts2}), only the
first (quartic) nonlinear correction to the $N=4$ Maxwell action
is known \cite{Tseyt}.

Because of lacking the Goldstone superfield BI actions
for these cases, it is natural to try to deduce
the relevant superfield equations of motion as a non-polynomial
generalization of the $N=2$ and $N=4$ super
Maxwell equations. In \cite{BIK3,IKLe} we derived covariant superfield
equations for a few examples of superbranes
in $D=3$ and $D=4$ by applying the formalism of nonlinear
realizations to the appropriate partially broken supersymmetries.
In particular, we recovered the $N=2 \rightarrow N=1$ BI
system in this new setting \cite{BIK3}.

Here we apply the same method to derive the superfield
equations for the partial-breaking patterns
$N=4 \rightarrow  N=2$ and $N=8 \rightarrow N=4$ in $D=4$,
with the $N=2$ and $N=4$ vector multiplets as the Goldstone ones.
In both cases, the covariant gauge field strength satisfies a
disguised form of the BI equation which takes the familiar form
after a field redefinition. We elaborate in more detail on the
first case. In particular, we show
that the equation for the complex scalar field corresponds to
the static-gauge Nambu-Goto action of 3-brane in $D=6$.
To first orders in the Goldstone superfield, we find the
equivalence transformation to the ordinary $N=2$ vector multiplet
superfield strength and restore the
invariant off-shell action to sixth order. The first
correction to the action of \cite{Ket} arises in sixth order and
it coincides with the one found in \cite{SKT} from a different
reasoning. We speculate on a possibility to generate
non-abelian versions of super BI theories from the nonlinear-realizations 
approach.

\section{Vector Goldstone multiplet for $N=4 \rightarrow N=2$}

We wish to derive the $N=2$ supersymmetric BI theory as a theory
of the partial breaking of $N=4, D=4$ supersymmetry down to $N=2$
supersymmetry, with the vector $N=2$ multiplet as the Goldstone one. To
apply the nonlinear realizations techniques \cite{BIK3}, we firstly 
need to specify $N=4, D=4$ supersymmetry to start with. 
In the case of $N=2 \rightarrow N=1$
breaking
the basic object was the Goldstone $N=1$ spinor superfield associated 
with the
spontaneously broken half of the $N=2$ fermionic generators.
This superfield is related by a field redefinition to the standard
chiral spinor $N=1$ Maxwell superfield strength.
In the $N=2$ Maxwell theory, the basic object is a complex scalar
$N=2$ off-shell superfield strength ${\cal W}$ which is chiral and
satisfies one additional Bianchi identity:
\be
(a)\;\bar D^i_{\dot\alpha} {\cal W} =0~,\;
D_i^{\alpha} \bar{\cal W} =0~,\quad (b)\;
D^{ik}{\cal W} = \bar D^{ik}\bar{\cal  W}~. \lb{oshN2M}
\ee
Here,
\bea
&&D_{\alpha}^i = \frac{\partial}{\partial\theta^{\alpha}_i}+
  i{\bar\theta}^{{\dot\alpha}i}\partial_{\alpha\dot\alpha}\; ,\quad
{\bar D}_{{\dot\alpha}i}=-\frac{\partial}
{\partial{\bar\theta}^{\dot\alpha i}}
  -i\theta^{\alpha}_i\partial_{\alpha\dot\alpha}\;, \quad
 \left\{ D_{\alpha}^i, {\bar D}_{{\dot\alpha}j}\right\}=
 -2i\delta^i_j \partial_{\alpha\dot\alpha}~,\label{semicd} \\
&& D^{ij} \equiv  D^{\alpha\, i}D_{\alpha}^j~, \quad
\bar D^{ij} \equiv  \bar D^{i}_{\dot\alpha}D^{\dot\alpha \,j}~. \lb{defDD}
\eea
The superfield equation of motion for ${\cal W}$ reads
\be
D^{ik}{\cal W} + \bar D^{ik}\bar{\cal W} = 0~, \lb{eqm1}
\ee
and, together with (\ref{oshN2M}$b$), amounts to
\be
D^{ik}{\cal W}  = \bar D^{ik}\bar{\cal W} = 0~. \lb{eqmN2M}
\ee

In order to incorporate an appropriate generalization of 
${\cal W}$ into the nonlinear realization scheme as the 
Goldstone superfield, we need to have
the proper bosonic generator in the algebra. The following central 
extension of $N=4, D=4$ Poincar\'e
superalgebra suits this purpose
\bea
&& \left\{ Q_{\alpha}^i, {\bar Q}_{\dot{\alpha}j}
        \right\}=2\delta^i_jP_{\alpha\dot{\alpha}} \;, \quad
\left\{ S_{\alpha}^i, {\bar S}_{\dot{\alpha}j}
        \right\}=2\delta^i_jP_{\alpha\dot{\alpha}}\;, \nn
&& \left\{ Q_{\alpha}^i,  S_{\beta}^j
   \right\}=2\varepsilon^{ij}\varepsilon_{\alpha\beta}Z \;, \quad
      \left\{ {\bar Q}_{\dot{\alpha}i}, {\bar S}_{\dot{\beta}j}
        \right\}=
 -2\varepsilon_{ij}\varepsilon_{\dot{\alpha}\dot{\beta}}{\bar Z} \;,
\;\; (i, j = 1,2)~, \label{n4sa}
\eea
with all other (anti)commutators vanishing. Note an important feature
that the complex central charge $Z$ appears in the crossing
anticommutator, while the generators $(Q, \bar Q)$ and $(S, \bar S)$
on their own form two $N=2$ superalgebras without central charges.
The full internal symmetry automorphism
group of \p{n4sa} (commuting with $P_{\alpha\dot\alpha}$ and $Z$)
is $SO(5)\sim Sp(2)$. Besides the manifest $R$-symmetry
group $U_R(2) = SU_{R}(2)\times U_R(1)$
acting as uniform rotations of the doublet indices of all spinor
generators and the opposite phase transformations of the $S$- and
$Q$-generators, it also includes the 6-parameter quotient
$SO(5)/U_{R}(2)$ transformations
which properly rotate the generators $Q$ and $S$ through each other. The
superalgebra \p{n4sa} is a $D=4$ form of
$N=(2,0)$ (or $N=(0,2)$) Poincar\'e superalgebra in $D=6$.

As a first step towards the corresponding nonlinear realization,
let us split the set of generators of the $N=4$ superalgebra \p{n4sa}
into the unbroken $\left\{ Q_{\alpha}^i,{\bar Q}_{\dot\alpha j},
P_{\alpha\dot\alpha} \right\}$ and broken $\left\{ S_{\alpha}^i,
{\bar S}_{\dot\alpha j}, Z,{\bar Z} \right\}$ parts. A coset element
$g$ is then defined by:
\be\label{vectorcoset}
g= \mbox{exp}\,i\left(-x^{\alpha\dot{\alpha}}P_{\alpha\dot{\alpha}} +
\theta^{\alpha}_i Q^i_{\alpha}+{\bar\theta}^i_{\dot\alpha}
    {\bar Q}_i^{\dot\alpha}\right)
  \mbox{exp}\,i\left(\psi^{\alpha}_i S^i_{\alpha}+{\bar\psi}^i_{\dot\alpha}
    {\bar S}_i^{\dot\alpha}\right)
   \mbox{exp}\,i\left(WZ+{\bar W}{\bar Z}\right) \;.
\ee
Acting on \p{vectorcoset} from the left by various elements
of the supergroup corresponding to \p{n4sa}, one can find the
transformation properties of the coset coordinates.

For the unbroken supersymmetry $\left( g_0=\mbox{exp}\,i
\left( -a^{\alpha\dot\alpha}P_{\alpha\dot\alpha}+
\epsilon^{\alpha}_i Q^i_{\alpha}+{\bar\epsilon}^i_{\dot\alpha}
    {\bar Q}_i^{\dot\alpha}\right)\right) $ one has:
\be\label{unbrokentr}
\delta x^{\alpha\dot\alpha}=a^{\alpha\dot\alpha} -i\left(
\epsilon^{\alpha}_i{\bar\theta}^{\dot\alpha i}+
{\bar\epsilon}^{\dot\alpha i}\theta^{\alpha}_i\right) \;, \quad
\delta\theta^{\alpha}_i=\epsilon^{\alpha}_i\; , \quad
\delta{\bar\theta}^i_{\dot\alpha}=\bar\epsilon^i_{\dot\alpha} \;.
\ee

Broken supersymmetry transformations
$\left( g_0=\mbox{exp}\,i
\left(
\eta^{\alpha}_i S^i_{\alpha}+{\bar\eta}^i_{\dot\alpha}
    {\bar S}_i^{\dot\alpha}\right)\right) $ are as follows:
\bea\label{brokentr}
&&\delta x^{\alpha\dot\alpha}= -i\left(
\eta^{\alpha}_i{\bar\psi}^{\dot\alpha i}+
{\bar\eta}^{\dot\alpha i}\psi^{\alpha}_i\right) \;, \quad
\delta\psi^{\alpha}_i=\eta^{\alpha}_i\; , \quad
\delta{\bar\psi}^i_{\dot\alpha}=\bar\eta^i_{\dot\alpha} \;, \nn
&&\delta W = -2i\eta^{\alpha}_i\theta^i_{\alpha} \:, \quad
\delta{\bar W}=-2i{\bar\eta}^i_{\dot\alpha}{\bar\theta}_i^{\dot\alpha}\;.
\eea

Finally, the
broken $Z,{\bar Z}$-translations $\left( g_0=\mbox{exp}\,i
\left( cZ+{\bar c}{\bar Z}\right)\right) $ read
\be\label{brokenZ}
\delta W = c \:, \quad
\delta{\bar W}= {\bar c}\;.
\ee

The next standard step is to define
the left-invariant Cartan 1-forms:
\bea\label{vectorcf}
\omega_P^{\alpha\dot\alpha} &=& dx^{\alpha\dot\alpha}-
  i\left( d{\bar\theta}^{{\dot\alpha}i} \theta_i^{\alpha} +
  d\theta_i^{\alpha}{\bar\theta}^{{\dot\alpha}i}\right) -i
  \left(d{\bar\psi}^{{\dot\alpha}i} \psi_i^{\alpha}+
  d\psi_i^{\alpha}{\bar\psi}^{{\dot\alpha}i}\right)~, \nn
\omega_{Q\; i}^{\alpha} &=& d\theta^\alpha_i \; , \quad
   {\bar\omega}_{Q\;\dot\alpha}^{i}= d{\bar\theta}_{\dot\alpha}^{i} \; , 
\quad
   \omega_{S\; i}^{\alpha} = d\psi^\alpha_i \; , \quad
   {\bar\omega}_{S\;\dot\alpha}^{i}= d{\bar\psi}_{\dot\alpha}^{i} \; , \nn
\omega_Z &=& dW-2i d\theta^\alpha_i  \psi^i_\alpha \; , \quad
   {\bar\omega}_Z=d{\bar W} +2i
        d{\bar\theta}^{{\dot\alpha}i} {\bar\psi}_{{\dot\alpha}i} \; .
\eea
The covariant derivatives of some scalar $N=2$ superfield $\Phi $ 
are defined by expanding
the differential $d\Phi$ over the covariant differentials of 
the $N=2$ superspace coordinates
\bea
&& d\Phi \equiv {\omega}_P^{\alpha\dot\alpha} 
\nabla_{\alpha\dot\alpha} \Phi+
 d\theta^\alpha_i \cD_\alpha^i \Phi +
   d\bar\theta_{\dot\alpha}^i{\cDb}^{\dot\alpha}_i \Phi \;\;\Rightarrow \nn
&& \nabla_{\alpha\dot\alpha}=
  \left( E^{-1} \right)_{\alpha\dot\alpha}^{\beta\dot\beta}
  \partial_{\beta\dot\beta}\; , \quad
E_{\alpha\dot\alpha}^{\beta\dot\beta}\equiv
\delta_{\alpha}^{\beta}\delta_{\dot\alpha}^{\dot\beta} +i
 \psi_i^{\beta}\partial_{\alpha\dot\alpha}{\bar\psi}^{\dot\beta i}
 +i {\bar\psi}^{\dot\beta i}\partial_{\alpha\dot\alpha}
 \psi^\beta_i \;, \nn
&& \cD_{\alpha}^i = D_\alpha^i +i \left( \psi^\beta_j D_{\alpha}^i
  {\bar\psi}^{\dot\beta j}+{\bar\psi}^{\dot\beta j}D_{\alpha}^i
      \psi^\beta_j \right)\nabla_{\beta\dot\beta} \; , \nn
&& \cDb_{\dot\alpha i} = {\bar D}_{\dot\alpha i} +i \left( \psi^\beta_j
 {\bar D}_{\dot\alpha i} {\bar\psi}^{\dot\beta j}+{\bar\psi}^{\dot\beta j}
 {\bar D}_{\dot\alpha i}  \psi^\beta_j \right)\nabla_{\beta\dot\beta} \; ,
\label{fullcd}
\eea
where $D^i_\alpha, \bar D_{i\dot\alpha}$ are defined in \p{semicd}.
The derivatives \p{fullcd} obey the following algebra:
\bea\label{algebracovd}
&& \left\{ \cD^i_{\alpha}, \cDb_{ {\dot\alpha}j}\right\}=
 -2i\delta^i_j \nabla_{\alpha\dot\alpha} +2i\left(
 \cD^i_{\alpha}\psi_k^{\gamma}\cDb_{\dot\alpha j}{\bar\psi}^{\dot\gamma k}+
 \cD^i_{\alpha}{\bar\psi}^{\dot\gamma k}\cDb_{\dot\alpha j}\psi^{\gamma}_k
 \right) \nabla_{\gamma\dot\gamma}\; , \nn
&&\left\{ \cD^i_{\alpha}, \cD^j_{\beta}\right\}=
  2i\left(
 \cD^i_{\alpha}\psi_k^{\gamma}\cD^j_{\beta}{\bar\psi}^{\dot\gamma k}+
 \cD^i_{\alpha}{\bar\psi}^{\dot\gamma k}\cD^j_{\beta}\psi^{\gamma}_k
 \right) \nabla_{\gamma\dot\gamma}\; , \nn
&&\left[ \cD^i_{\alpha}, \nabla_{ \beta\dot\beta}\right]=
 -2i\left(
 \cD^i_{\alpha}\psi_k^{\gamma}\cD_{\beta\dot\beta}{\bar\psi}^{\dot\gamma k}+
 \cD^i_{\alpha}{\bar\psi}^{\dot\gamma k}\cD_{\beta\dot\beta}\psi^{\gamma}_k
 \right) \nabla_{\gamma\dot\gamma}\, .
\eea

As in several previously studied examples \cite{BG0,BIK1,BIK2,IK1}, 
the Goldstone fermionic superfields
$\psi^i_\alpha$, ${\bar\psi}_{\dot\alpha i}$ can be covariantly
expressed in terms of the central-charge Goldstone superfields
${\cal W}$, $\bar{\cal W}$ by imposing the inverse Higgs
constraints \cite{IH} on the central-charge Cartan 1-forms. In the present
case these constraints are
\be
\omega_Z\vert_{d\theta, d\bar\theta} =
\bar\omega_Z\vert_{d\theta, d\bar\theta} = 0~,  \lb{IH}
\ee
where $\vert $ means the covariant projections on the differentials of
the spinor coordinates. These constraints amount to the sought expressions
for the fermionic Goldstone superfields
\be\label{vectorih}
\psi^i_\alpha = -\frac{i}{2}\cD_{\alpha}^i W\;,\quad
{\bar\psi}_{\dot\alpha i}=-\frac{i}{2}\cDb_{\dot\alpha i}{\bar W}~,
\ee
and, simultaneously, to the covariantization of the chirality conditions
(\ref{oshN2M}$a$)
\be
\cDb_{\dot\alpha i} W =0 \; , \quad \cD_{\alpha}^i {\bar W}=0 \;.
 \label{vectorchir}
\ee
Actually, eqs. \p{vectorih} are highly nonlinear equations serving to
express $\psi^i_\alpha$, ${\bar\psi}_{\dot\alpha i}$
in terms of $W, \bar W$ with making use of the 
definitions \p{fullcd}.

It is also straightforward to write the covariant generalization of the
dynamical equation of the $N=2$ abelian vector multiplet \p{oshN2M},
\p{eqmN2M}
\bea
\cD^{\alpha( i}\cD_{\alpha}^{j)} W= 0\;,\quad
   \cDb_{\dot\alpha}^{(i} \cDb^{\dot\alpha j)}{\bar W} =0 \; .
  \label{vectoreom}
\eea

The equations \p{vectorchir}, \p{vectoreom} with the superfield 
Goldstone fermions
eliminated by \p{vectorih} constitute a manifestly covariant form of the
superfield equations of motion of $N=2$ Dirac-Born-Infeld theory with
the second hidden nonlinearly realized $N=2$ supersymmetry. It closes, 
together with the manifest $N=2$ supersymmetry, 
on the $N=4$ supersymmetry \p{n4sa}.

As a first step in proving this statement, let us show that the above system
of equations reduces the component content of $W$ just to that of 
the on-shell
$N=2$ vector multiplet. It is convenient to count the number of independent
covariant superfield projections of $W$, $\bar W$.

At the dimensions $(-1)$ and $(-1/2)$ we find $W$, $\bar W$ and
$\psi_{i\alpha} = -\frac{i}{2}\cD_{i\alpha} W\;,\quad
{\bar\psi}_{\dot\alpha}^i = \overline{(\psi_{i\alpha})}
=-\frac{i}{2}\cDb_{\dot\alpha}^i{\bar W}$, with a complex bosonic field and
a doublet of gaugini as the lowest components.

At the dimension $(0)$ we have, 
before employing \p{vectorchir}, \p{vectoreom},
\bea
&& \cD_{\alpha}^i\psi_{\beta}^j=
 \varepsilon^{ij}f_{\alpha\beta} +
i\varepsilon_{\alpha\beta}F^{(ij)} +
F^{(ij)}_{(\alpha\beta)}\; , \quad
 \cDb_{\dot\alpha i}\psi_{j\alpha} =
\varepsilon_{ij} X_{\alpha\dot\alpha} + X_{(ij)\alpha\dot\alpha}\;, \nn
&& \cDb_{\dot\alpha i}{\bar \psi}_{\dot\beta j} =
 -\varepsilon_{ij}\bar f_{\dot\alpha \dot\beta} +i
\varepsilon_{\dot\alpha\dot\beta}\bar F_{(ij)} +
\bar F_{(ij)(\dot\alpha\dot\beta)}\; \quad
 \cD_\alpha^i{\bar \psi}_{\dot\alpha}^j = \varepsilon^{ij}
   {\bar X}_{\dot\alpha\alpha}-\bar X^{(ij)}_{\dot\alpha\alpha} \;, 
\label{components11} \\
&&f_{\alpha\beta} \equiv \epsilon_{\alpha\beta}A + iF_{\alpha\beta},
\; \bar f_{\dot\alpha\dot\beta} = \overline{(f_{\alpha\beta})} =
\epsilon_{\dot\alpha\dot\beta}\bar A - i\bar F_{\dot\alpha\dot\beta},
\\
&& f^\alpha_\beta f_{\alpha\gamma} = \epsilon_{\beta\gamma}\left(A^2 -
{1\over 2}F^2\right)~, \quad
\bar f^{\dot\alpha}_{\dot\beta} \bar f_{\dot\alpha\dot\gamma} =
\epsilon_{\dot\beta\dot\gamma}\left(\bar A^2 - {1\over 2}\bar F^2\right)~.
\eea

The dynamical equations \p{vectoreom} imply
\be
F^{(ij)} = \bar F^{(ij)} = 0~. \label{aux}
\ee
The lowest component of these superfields is a nonlinear analog
of the auxiliary field of the $N=2$ Maxwell theory.

Next, substituting the expressions \p{vectorih} for the spinor
Goldstone fermions in the l.h.s. of eqs. \p{components11} and making use
of both \p{vectorchir} and \p{vectoreom}, we represent these l.h.s. as
\bea
&& \cD_{\alpha}^i\psi_{\beta}^j=
-\frac{i}{4}\left\{ \cD_{\alpha}^i,\cD_{\beta}^j\right\}W +
 \frac{i}{4}\varepsilon^{ij} \cD_{( \alpha}^k\cD_{\beta ) k}W \; , \;
 \cDb_{\dot\alpha i}\psi_\alpha^j = -\frac{i}{2}
 \left\{ \cDb_{\dot\alpha i} , \cD_{\alpha}^j \right\} W \;, \nn
&& \cDb_{\dot\alpha i}{\bar \psi}_{\dot\beta j} =
-\frac{i}{4}\left\{ \cDb_{\dot\alpha i},\cDb_{\dot \beta j}\right\}{\bar W} 
- \frac{i}{4}\varepsilon_{ij} \cDb_{( \dot\alpha}^k\cDb_{\dot\beta ) k}
   {\bar W} \; , \;
 \cD_\alpha^i{\bar \psi}_{\dot\alpha j} =
 -\frac{i}{2}
 \left\{ \cD_{\alpha}^i , \cDb_{\dot\alpha j} \right\}
  {\bar W} \;. \label{components2}
\eea
Using the algebra \p{algebracovd} and comparing \p{components2}
with the definition \p{components11} (taking into account \p{aux}),
it is straightforward to show that the objects $F^{(ij)}_{(\alpha\beta)}$,
$\bar F_{(ij)(\dot\alpha\dot\beta)}$, $X_{(ij)\alpha\dot\alpha}$
and $\bar X^{(ij)}_{\dot\alpha\alpha}$ satisfy a system
of homogeneous equations, such that the matrix of the coefficients
in them is nonsingular at the origin $W=\bar W = 0$. Thus these objects
vanish as a consequence of the basic equations:
\be
F^{(ij)}_{(\alpha\beta)} = \bar F_{(ij)(\dot\alpha\dot\beta)} 
= X_{(ij)\alpha\dot\alpha}
= \bar X^{(ij)}_{\dot\alpha\alpha} = 0. \lb{vantens}
\ee

As a result, on shell we are left with the following superfield content:
\bea
&& \cD_{\alpha}^i\psi_{\beta}^j=
 \varepsilon^{ij}f_{\alpha\beta}\; , \quad
 \cDb_{\dot\alpha i}\psi_{j\alpha} =
\varepsilon_{ij} X_{\alpha\dot\alpha}\;, \nn
&& \cDb_{\dot\alpha i}{\bar \psi}_{\dot\beta j} =
 -\varepsilon_{ij}\bar f_{\dot\alpha \dot\beta}\;, \quad
 \cD_\alpha^i{\bar \psi}_{\dot\alpha}^j = \varepsilon^{ij}
   {\bar X}_{\dot\alpha\alpha} \;. \label{components1}
\eea The only new independent superfield at the dimension $(0)$ is the
complex one $F_{(\alpha \beta)}, \bar F_{(\dot\alpha\dot\beta)}$, while $A,
\bar A$ and $X_{\alpha\dot\beta}, \bar X_{\dot\alpha\beta}$ are
algebraically expressed through it and other independent superfields
as will be shown below. In Sec.
3 we show that this superfield is related, by an equivalence
field redefinition, to the Maxwell field strength obeying 
the BI equation of motion.

Substituting the explicit expressions for the anticommutators 
\p{algebracovd}
into \p{components2} and again using \p{components1} in both sides 
of \p{components2},
we finally obtain:
\bea
&& A =
-{1\over 2} \bar X^{\;\;\beta}_{\dot\gamma}f_{\beta\gamma}
{\nabla}^{\gamma\dot\gamma}W~, \quad \bar A = -{1\over 2}
X^{\;\;\dot\beta}_{\gamma} \bar
f_{\dot\beta\dot\gamma}\nabla^{\gamma\dot\gamma}\bar W~, \nn
&&
X_{\alpha\dot\alpha } = {\nabla}_{\alpha\dot\alpha}W+ \left(
X_{\gamma\dot\alpha}{\bar X}_{\dot\gamma\alpha} + \bar f_{\dot\alpha
\dot\gamma}f_{\alpha\gamma}\right)
{\nabla}^{\gamma\dot\gamma} W \; ,\nn
&& {\bar X}_{\dot\alpha\alpha} = {\nabla}_{\alpha\dot\alpha}{\bar W}+
\left(
X_{\gamma\dot\alpha}{\bar X}_{\dot\gamma\alpha}+
\bar f_{\dot\alpha \dot\gamma}f_{\alpha\gamma}\right)
{\nabla}^{\gamma\dot\gamma} {\bar W} \; .
\label{sys1}
\eea
It is easy to see that these algebraic equations indeed allow one to
express $A, \bar A$ and $X_{\alpha\dot\beta}, \bar X_{\alpha\dot\beta}$
in terms of $F_{(\alpha \beta)}$, $\bar F_{(\dot\alpha\dot\beta)}$ and 
${\nabla}_{\alpha\dot\alpha}W $, ${\nabla}_{\alpha\dot\alpha}\bar W$:
\be
A = {i\over 2}\, F^{(\beta}_{\;\;\gamma)}{\nabla}_{\beta\dot\gamma}\bar W 
{\nabla}^{\gamma\dot\gamma}
W + \ldots~, \quad X_{\alpha\dot\alpha} = {\nabla}_{\alpha\dot\alpha}W
+{1\over 2} ({\nabla}W\cdot {\nabla}W) {\nabla}_{\alpha\dot\alpha}\bar W
+ \ldots ~.
\ee

As one more corollary of eqs \p{sys1}, let us check
the validity of additional integrability conditions which
come out from our covariant chirality constraints \p{vectorchir}
\be
\cD_{\alpha}^{ i}{\bar W}=\cDb_{\dot\alpha i}W=0 \quad \Rightarrow \quad
\left\{ \cD_{\alpha}^{ i},\cD_{\beta}^{ j}\right\}{\bar W}=
\left\{ \cDb_{ \dot\alpha i},\cDb_{\dot\beta j}\right\}W=0 \;. \lb{integr}
\ee
In terms of the components \p{components1} they read as follows:
\be
Y \equiv
\bar X^{\;\;\beta}_{\dot\beta}
f_{\beta\gamma}{\nabla}^{\gamma\dot\beta}\bar W =0~, \quad \bar Y
\equiv  X^{\;\;\dot\beta}_{\beta}
\bar f_{\dot\beta\dot\gamma}{\nabla}^{\dot\gamma\beta} W =0~.
\label{sys2}
\ee
Substituting  the expression for $\bar X_{\dot\beta\beta}$
from \p{sys1} into $Y$, we find after some algebra
\be \label{1}
\left[1 - \left(X\cdot{\nabla} \bar W\right)\right] Y = 
\left(\nabla\bar
W\right)^2 \left[\bar A\left({1\over 2}F^2 - A^2\right) - A + 
{1\over 2} \bar X^{\dot\gamma\beta}f_{\beta\lambda}
X^\lambda_{\;\;\dot\gamma}\right] \equiv
\left(\nabla\bar W\right)^2 {\cal B}~.
\ee
So, in order to prove \p{sys2} and \p{integr}, it suffices to show that
\be \label{2}
{\cal B} =\bar{\cal B} = 0~.
\ee

After some algebraic manipulations one gets
\be \label{4}
\left[1 - \left(\bar X\cdot\nabla W\right) + 
{1\over 4}\bar X^2 \left(\nabla 
W\right)^2 \right] {\cal B} =
\left(A^2 - {1\over 2}F^2\right)\left[\left(\nabla W
\cdot\nabla \bar W\right)\bar{\cal B} +
{1\over 2} \left(\bar X\cdot\nabla W\right)\bar
Y\right]~.
\ee
Recalling that $Y, \bar Y$ are expressed through
${\cal B}, \bar{\cal B}$ by division by non-singular factors 
(see eq. \p{1}),
and substituting these expressions into \p{4} and its conjugate,
we get for ${\cal B}, \bar{\cal B}$ a system of homogeneous linear
equations with a non-singular matrix of the coefficients.
This proves \p{2} and \p{sys2}.

Returning to the issue of extracting an irreducible set of covariant
superfield projections of $W, \bar W$, it is easy to show that the further
successive action by covariant spinor derivatives on \p{components1} 
produces no new independent superfields. One obtains either 
the equations of motion
(and Bianchi identities) for the independent basic superfields $W, \bar W,
\psi_{i\alpha}, \bar\psi^i_{\dot\alpha}$ and $F_{(\alpha\beta)}, \bar
F_{(\dot\alpha\dot\beta)}$, or some composite superfields 
which are expressed through $x$-derivatives of the basic ones 
(or as some appropriate nonlinear functions
of the basic superfields). The useful relations which essentially
simplify the analysis are the following ones:
\be
\left\{ \cD^{(i}_{\alpha}, \cDb^{j)}_{ {\dot\alpha}}\right\} =
\left\{ \cD^{(i}_{\alpha}, \cD^{j)}_{\beta}\right\} =
\left\{ \cDb^{(i}_{\dot\alpha}, \cDb^{j)}_{\dot\alpha}\right\}
= 0~. \lb{harmcov}
\ee
They follow by substituting \p{components1} into the algebra
\p{algebracovd}. These relations are the covariant version of the 
integrability
conditions for the Grassmann harmonic
$N=2$ analyticity \cite{Harm}. Thus the nonlinear $W, \bar W$
background specified by the equations \p{vectorih}-\p{vectoreom} respects
the Grassmann harmonic analyticity which plays a fundamental role
in $N=2, D=4$ theories.

Before going further, let us make a few comments.

First, the nonlinear realization setting we used,
in order to deduce our equations
\p{vectorih}-\p{vectoreom}, drastically differs from the standard superspace
differential-geometry setup of supersymmetric gauge theories
(see, e.g., \cite{bw,Sohn}). The starting
point of the standard approach is the covariantization of the flat
derivatives (spinor and vector) by the gauge-algebra valued connections with
appropriate constraints on the relevant covariant superfield
strengths. In our case (quite analogously to the previously considered $N=1,
D=4$ \cite{BG2} and $N=1, D=3$ \cite{IK1} cases) the covariant 
derivatives include
no connection-type terms. Instead, they contain, in a highly non-linear
manner, the Goldstone bosonic $N=2$ superfields $W, \bar W$. These
quantities, after submitting them to the covariant constraints
\p{vectorih}-\p{vectoreom}, turn out to be the nonlinear-realization
counterparts of the $N=2$ Maxwell superfield strength. As we show below, 
the Bianchi identities needed to pass to the gauge potentials are 
encoded in the set \p{vectorih}-\p{vectoreom}.

In the differential-geometry approach the
constraints like \p{vantens} emerge before going on shell,
they are a consequence of the Bianchi identities.
In our nonlinear system we cannot separate in a simple way the kinematical
off-shell constraints from the dynamical on-shell ones. We could try to
relax our system by lifting the basic dynamical equations \p{vectoreom}
and retaining only the chirality condition \p{vectorchir} together
with \p{vectorih} and an appropriate covariantization of
the constraint (\ref{oshN2M}$b$). But in this case we
immediately face the same difficulty as in the
$N=2 \rightarrow N=1$ case \cite{BG2}:
a naive covariantization of (\ref{oshN2M}$b$) by replacing the flat
spinor derivatives by the
covariant ones proves to be not self-consistent. For self-consistency, it
should be properly modified order by order, without any clear guiding
principle. No such a problem
arises when the dynamical equations \p{vectoreom} are enforced. The
terms modifying the naive covariantization of (\ref{oshN2M}$b$) 
can be shown to
vanish, as in the $N=2 \rightarrow N=1$ case \cite{BIK3}.

Nevertheless, as we argue below, there exists a highly nonlinear
field redefinition which relates the nonlinear superfield
Goldstone strength $W, \bar W$ to its flat counterpart
${\cal W}, \bar{\cal W}$
satisfying the off-shell irreducibility
conditions (\ref{oshN2M}). In this frame it becomes possible 
to divide the kinematical and
dynamical aspects of our system and to write the appropriate off-shell
action giving rise to the dynamical equations, in a deep analogy
with the $N=2 \rightarrow N=1$ case \cite{BG2,RT}.

As the last comment, we note that all the fields of the multiplet comprised
by $W, \bar W$, except for $F_{(\alpha\beta)}, \bar
F_{(\dot\alpha\dot\beta)}$, can be given a clear interpretation as
Goldstone fields: $W \vert, \bar W \vert$ for the spontaneously broken
central-charge shifts, $\psi_{\alpha}^i\vert, \bar\psi_{i\dot\alpha}\vert$
for the spontaneously broken $S$-supersymmetry transformations and $F^{(ij)}
\vert, \bar F^{(ij)} \vert$ for the spontaneously broken $SO(5)/U_{R}(2)$
transformations. This immediately follows from considering the
transformation properties of the coset element \p{vectorcoset}. Eq. \p{aux}
following from the dynamical equations \p{vectoreom} explicitly breaks the
$SO(5)/U_{R}(2)$ symmetry, leaving us with $U_{R}(2)\times U(1)$ as the only
surviving internal symmetry (an extra $U(1)$ factor realized as a phase on
$W, \bar W$ and spinor $N=2$ superspace coordinates comes from the $D=6$
Lorentz group). This is quite similar to the $N=2 \rightarrow N=1$ cases
\cite{BG0,BG2}, where the $U(2)$ automorphism symmetry of the original 
$N=2$ Poincar\'e superalgebra proves to be finally broken down to some
its subgroup.

\setcounter{equation}{0}
\section{Bosonic equations of motion}
As the next important step in examining the superfield system
\p{vectorih}-\p{vectoreom}, we inspect its bosonic sector. The set of
bosonic equations can be obtained by acting on both sides of \p{components1}
by two covariantized spinor derivatives, using the algebra \p{algebracovd}
together with the relations \p{sys1} and omitting the fermions in the final
expressions (which should contain only independent superfield projections
and their $x$-derivatives). Instead of analyzing the bosonic sector in full
generality, we specialize here to its two suggestive limits.

\vspace{0.2cm}
\noindent{\it 1. Vector fields limit.}
This limit amounts to
\be\label{1case}
 W \left|_{\theta=\bar\theta=0} =
 {\bar W}\right|_{\theta=\bar\theta=0} = 0 \; .
\ee
{}From eqs. \p{sys1} with all fermions omitted, one can see that
\p{1case} imply
\be\label{1casea}
A={\bar A}=X_{\alpha\dot\alpha}={\bar X}_{\dot\alpha\alpha}=0 .
\ee
Thus, in this limit our superfields $W,{\bar W}$ contain only
$F_{\alpha\beta}, {\bar F}_{\dot\alpha \dot\beta}$ as the bosonic components,
which, owing to \p{vectoreom}, obey the following simple equations
\be\label{case1eom}
\partial_{\alpha\dot\alpha} F^{\alpha\beta} -F_\alpha^\gamma
  {\bar F}_{\dot\alpha}^{\dot\gamma}\partial_{\gamma\dot\gamma}
F^{\alpha\beta}=0 \;, \quad
\partial_{\alpha\dot\alpha} {\bar F}^{\dot\alpha \dot\beta} 
-F_\alpha^\gamma
  {\bar F}_{\dot\alpha}^{\dot\gamma}\partial_{\gamma\dot\gamma}
{\bar F}^{\dot\alpha \dot\beta}=0\;.
\ee
It was already shown in \cite{BIK3} that one can  split eqs.
\p{case1eom} into the ``true'' Bianchi identities and
``true'' equations of motion
\be\label{case1eom1}
\partial_{\beta\dot\alpha} \left( f F_{\alpha}^{\beta}\right) -
\partial_{\alpha\dot\beta}\left( {\bar f}{\bar F}_{\dot\alpha}^{\dot\beta}
  \right) =0 \;, \quad
\partial_{\beta\dot\alpha} \left( g F_{\alpha}^{\beta}\right) +
\partial_{\alpha\dot\beta}\left( {\bar g}{\bar F}_{\dot\alpha}^{\dot\beta}
  \right) =0 \;,
\ee
where
\be
f=\frac{ {\bar F}^2-2}{1-{1\over 4} F^2 {\bar F}^2}\;, \quad
g=\frac{ {\bar F}^2+2}{1-{1\over 4} F^2 {\bar F}^2}\;.
\ee
Now, in terms of the ``genuine'' field strengths
\be
V_\alpha^\beta \equiv \frac{1}{2\sqrt{2}}fF_\alpha^\beta \; , \quad
{\bar V}_{\dot\alpha}^{\dot\beta} \equiv
 \frac{1}{2\sqrt{2}}{\bar f}{\bar F}_{\dot\alpha}^{\dot\beta} \; ,
\ee
the first equation \p{case1eom1} coincides with the standard Bianchi
identity
\be
\partial_{\beta\dot\alpha}V_\alpha^\beta-\partial_{\alpha\dot\beta}
  {\bar V}_{\dot\alpha}^{\dot\beta} =0\; , \lb{bianchiF}
\ee
while the second one coincides with the equation of motion following 
from the $D=4$ BI action:
\be\label{BI1}
S=\int d^4x \sqrt{ \left( V^2-{\bar V}{}^2\right)^2 -2
  \left( V^2 +{\bar V}{}^2\right) +1 } \;.
\ee

Thus the superfield system \p{vectorih}-\p{vectoreom} encodes the
BI equation, in accord with the statement that this system provides
a supersymmetric extension of the latter.

Note that our original variables $F_{\alpha\beta},
{\bar F}_{\dot\alpha \dot\beta}$ allow
to avoid a square root  in \p{BI1}
\be
 \sqrt{ \left( V^2-{\bar V}{}^2\right)^2 -2
  \left( V^2 +{\bar V}{}^2\right) +1 } =
\frac{ \left( 1-\frac{1}{2} F^2 \right)\left( 1-\frac{1}{2}{\bar F}{}^2
 \right) }{1-\frac{1}{4} F^2{\bar F}{}^2} \; .
\ee
However, the Bianchi identity \p{bianchiF} becomes very complicated in
such a parametrization and it is unclear whether one can solve it in terms
of an appropriate vector potential.

\vspace{0.2cm}
\noindent{\it 2. Scalar fields limit.} This limit corresponds to 
the reduction
\be\label{2case}
 \cD^i_{(\alpha} \cD_{\beta) i}W \left|_{\theta=\bar\theta=0} =
 \cDb^i_{(\dot\alpha}\cDb_{\dot\beta) i}
{\bar W}\right|_{\theta=\bar\theta=0} = 0 \; .
\ee
{}From eqs. \p{sys1} and \p{sys2} one finds that the reduction
conditions \p{2case} imply
\bea\label{2casea}
&& A=0\; , \quad {\bar A}=0\;, \nn
&& X_{\alpha\dot\alpha}=\partial_{\alpha\dot\alpha}W +
  X_{\gamma\dot\alpha}{\bar X}_{\dot\gamma\alpha}
\partial^{\gamma\dot\gamma}W,
\quad
{\bar X}_{\dot\alpha\alpha}=\partial_{\alpha\dot\alpha}{\bar W} +
  X_{\gamma\dot\alpha}{\bar X}_{\dot\gamma\alpha}
 \partial^{\gamma\dot\gamma}{\bar W} ,
\eea
while the equations of motion following from  \p{vectoreom} read
\be\label{2caseb}
\partial_{\dot\alpha}^\alpha X_{\alpha\dot\beta} +
 {\bar X}^{\dot\gamma\alpha} X_{\;\;\dot\alpha}^\gamma
 \partial_{\gamma\dot\gamma} X_{\alpha\dot\beta}=0\; , \quad
\partial_{\alpha\dot\alpha} {\bar X}^{\dot\alpha}_{\;\;\beta} +
 {\bar X}_{\;\;\alpha}^{\dot\gamma} X_{\;\;\dot\alpha}^\gamma
 \partial_{\gamma\dot\gamma} {\bar X}_{\;\;\beta}^{\dot\alpha}=0\; .
\ee
The system \p{2casea} can be easily solved
\be\label{sol1}
X_{\alpha\dot\alpha}=\partial_{\alpha\dot\alpha} W +
 \frac{\left( \partial W\right)^2 }{h}\partial_{\alpha\dot\alpha}
 {\bar W}\;, \quad
{\bar X}_{\dot\alpha\alpha}=\partial_{\alpha\dot\alpha} {\bar W} +
 \frac{\left( \partial {\bar W}\right)^2 }{h}\partial_{\alpha\dot\alpha}
 { W}\;,
\ee
where
\be\label{sol1a}
h=\left[ 1- \left(\partial W\cdot
 \partial {\bar W}\right)\right] +
 \sqrt{ \left[ 1- \left(\partial W\cdot\partial {\bar W}\right)\right]^2 
- \left( \partial W\right)^2
 \left( \partial{\bar W}\right)^2} \;.
\ee
One can check that those parts of eqs. \p{2caseb} which are symmetric
in the free indices are identically satisfied with \p{sol1} and \p{sol1a}.
The trace part of \p{2caseb} can be cast into the form:
\be\label{scalareq1}
\partial_{\alpha\dot\alpha}\left( \frac{X^{\alpha\dot\alpha}+
 \frac{1}{2}X^2 {\bar X}^{\dot\alpha\alpha} }{1-\frac{1}{4}X^2{\bar X}{}^2}
 \right)=0 \;, \quad
\partial_{\alpha\dot\alpha}\left( \frac{{\bar X}^{\dot\alpha\alpha}+
 \frac{1}{2}{\bar X}{}^2  X^{\alpha\dot\alpha} }
{1-\frac{1}{4}X^2{\bar X}{}^2}
 \right)=0 \;.
\ee
Now, substituting \p{sol1}, \p{sol1a} in \p{scalareq1}, one finds
that the resulting form of these equations can be reproduced from the action
\be\label{NGaction} S=\int
d^4x\left( \sqrt{ 1+2\left(\partial W
\cdot \partial {\bar W}\right)+ \left(\partial W\cdot
 \partial {\bar W}\right)^2 -
 \left( \partial W\right)^2
 \left( \partial{\bar W}\right)^2}-1\right) \;.
\ee
This action is the static-gauge form of the Dirac-Nambu-Goto
action of a 3-brane in D=6.

In terms of the variables
$X_{\alpha\dot\alpha},{\bar X}_{\dot\alpha\alpha}$ the action
\p{NGaction} becomes a rational function:
\be\label{NGaction1}
S=\int d^4x \left(
\frac{ 1+ \left(X\cdot{\bar X}\right)+
\frac{1}{4} X^2 {\bar X}{}^2 }{ 1-\frac{1}{4} X^2 {\bar X}{}^2} -1
\right) \; .
\ee
Note that one cannot vary $X_{\alpha\dot\alpha}, \bar
X_{\dot\alpha\alpha}$ as independent fields, since, 
in view of \p{sol1}, they satisfy
some nonlinear integrability conditions.

Summarizing the discussion of Secs. 2 and 3, we have shown that the system
of our superfield equations \p{vectorih}-\p{vectoreom} is self-consistent
and gives a $N=2$ superextension of both the equations of $D=4$ BI theory
and those of the static-gauge 3-brane in $D=6$, with the nonlinearly
realized second $N=2$ supersymmetry. This justifies our claim that
\p{vectorih}-\p{vectoreom} are indeed a manifestly worldvolume 
supersymmetric form of the equations of D3-brane in $D=6$ and, 
simultaneously, of $N=2$ Born-Infeld theory. Similarly to the 
previous examples \cite{BIK3,IKLe}, the nonlinear realization approach 
yields the BI equations in a disguised form, with the Bianchi identities 
and dynamical equations mixed in a tricky way. At the same time, 
for the scalars we get the familiar static-gauge Nambu-Goto-type 
equations. This is
in agreement with the fact that $W, \bar W$ undergo pure shifts
under the action of the central charge generators $Z, \bar Z$, suggesting
the interpretation of these superfields as the transverse brane coordinates
conjugated to $Z, \bar Z$. These generators, in turn, can be
interpreted as two extra components of the 6-momentum.

\setcounter{equation}0
\section{Towards a formulation in terms of ${\cal W}, \bar{\cal W}$}

As was already mentioned, we expect that, like in the $N=1$
case \cite{BG2,RT}, there should exist an equivalence transformation 
to a formulation
in terms of the conventional $N=2$ Maxwell superfield strength 
${\cal W}, \bar{\cal W}$
defined by the off-shell constraints \p{oshN2M}.

A systematic, though as yet iterative procedure to
find such a field redefinition starts by passing to
the standard chirality conditions (\ref{oshN2M}$a$) from
the covariantly-chiral ones \p{vectorchir}. After some algebra,
\p{vectorchir} can be brought to the form
\be\label{chir2} {\bar D}{}_{\dot\alpha i} R=0 \;,
\quad D_{\alpha}^i {\bar R} =0 \;,
\ee
where
\bea\label{R}
R= W+\frac{1}{2}
 {\bar W}\left(\partial W\cdot \partial W\right) +\frac{i}{4}D^\gamma_j W
  {\bar D}^{\dot\gamma j}{\bar W}\partial_{\gamma\dot\gamma}W  + O(W^5)\; .
\eea
Now we pass to the new superfields ${\cal W},{\bar{\cal W}}$ with preserving
the flat chirality
\bea
\cW \equiv R\left(1-\frac{1}{2}{\bar D}{}^4{\bar R}{}^2\right)\;, \quad
\cWb \equiv {\bar R}\left(1-\frac{1}{2} D^4 R^2\right)\;, \quad
{\bar D}_{\dot\alpha i} \cW= D_\alpha^i \cWb =0 \;, \label{W}
\eea
where
\bea
D^4\equiv \frac{1}{48} D^{\alpha i}D^j_\alpha
D^{\beta}_{ i}D_{\beta j} \; , \quad \bar D^4 = 
\overline{(D^4)} \equiv\frac{1}{48}
\bar D_{\dot\alpha}^i \bar D^{\dot\alpha j}
   \bar D_{\dot\beta i}\bar D^{\dot\beta}_ j~. \label{definitions1}
\eea
Up to the considered third order, in terms of these superfields 
eqs. \p{vectoreom} can be rewritten as
\bea
&& D^{ij} \cW = {\bar D}{}^{ij}\cWb \; \Rightarrow \;
D^4 {\cal W} = -{1\over 2}\Box \bar{\cal W}~, \;
\bar D^4 \bar{\cal W} = -{1\over 2}\Box {\cal W}~, \; \Box \equiv
\partial_{\alpha\dot\alpha}\partial^{\alpha\dot\alpha}~, \label{bianchi} \\
&& D^{ij} \left( \cW + \cW\,{\bar D}{}^4{\cWb}^2 \right)+
{\bar D}{}^{ij} \left( \cWb +\cWb\, D^4{\cW}^2 \right) =0\;.\label{eom2}
\eea
Eq. \p{bianchi} is recognized as the Bianchi identity (\ref{oshN2M}$b$), so
${\cal W}, \bar{\cal W}$ can be identified with the conventional
$N=2$ vector multiplet superfield strength. Eq. \p{eom2} is then a
nonlinear generalization of the standard free $N=2$ vector multiplet
equation of motion \p{eqm1}. The transformation properties of $\cW,\cWb$
can be easily restored from \p{unbrokentr}-\p{brokenZ} and the 
definitions \p{R}, \p{W}.

The above procedure is an $N=2$ superfield analog of separating
Bianchi identities and dynamical equations for $F_{(\alpha\beta)},
\bar F_{(\dot\alpha\dot\beta)}$ (see Sec. 3). In both
cases we do not know the geometric principle behind the relevant field
redefinitions. Though in the bosonic case
we managed to find this redefinition in a closed form,
we are not aware of it in the full superfield case. Nonetheless, 
we can move a step further
and find the relation between
$W, \bar W$ and ${\cal W}, \bar{\cal W}$, as well as the
nonlinear dynamical equations for the latter,  up to the fifth order.
Then, using the transformation laws \p{unbrokentr}-\p{brokenZ}, we
can restore the hidden $S$-supersymmetry
and $Z, \bar Z$ transformations up to the fourth order.
In this approximation, the transformation laws and equations of motion read
\bea
&& \delta {\cal W} = f -{1\over 2}\bar D^4 (f A) +
{1\over 4} \Box (\bar f \bar A) + {1\over 4i}
\,\bar D^{i\dot\alpha}\bar f D^{\alpha}_i\,\partial_{\alpha\dot\alpha}\bar
A~, \quad \delta \bar{\cal W} = (\delta {\cal W})^*~,
\label{4transf} \\
&& A = \bar{\cal W}^2\left(1 + {1\over 2}D^4{\cal W}^2 \right)~, \quad
f = c+ 2i\,\eta^{i\alpha}\theta_{i\alpha}~, \quad \bar f =
\bar c  - 2i\,\bar\eta^i_{\dot\alpha} \bar\theta^{\dot\alpha}_i~,
\label{fdef} \\
&& D^{ij} B + \bar D^{ij} \bar B = 0~, \nn
&&B = {\cal W} +
{\cal W}\bar D^4 \left(\bar{\cal W}^2 + \bar{\cal W}^2D^4{\cal W}^2 +
{1\over 2}\bar{\cal W}^2 \bar D^4 \bar{\cal W}^2 - {1\over 6} {\cal W}
\Box \bar {\cal W}^3\right)~. \lb{5eq}
\eea
The hidden supersymmetry transformations,
up to the third order, close on the $c, \bar c$ ones in the $\eta,
\epsilon$ and $\bar\eta, \bar\epsilon$ sectors, and on the standard $D=4$
translations in the $\eta,  \bar\eta$ sector. In the sectors $\eta, \eta$
and $\bar\eta,  \bar\eta$ the transformations commute, as it should be.
Note that \p{4transf} is already of the most general form
compatible with the chirality conditions and Bianchi identity \p{oshN2M}. So
this form will be retained to any order, only the functions $A$,
$\bar A$ will get additional contributions.

It is straightforward to restore, to the sixth order, the off-shell
${\cal W}, \bar{\cal W}$ action which yields eq. \p{5eq} as the 
equation of motion
\bea
S^{(6)}_{bi} = {1\over 8}\left(\int d\zeta_L {\cal W}^2 
+ \mbox{c.c.}\right) +
{1\over 16}\int d Z \left\{ {\cal W}^2\bar{\cal W}^2\left[2 + \left(
D^4{\cal W}^2 + \bar D^4 \bar{\cal W}{}^2\right)\right] - 
{1\over 9} {\cal
W}^3 \Box \bar{\cal W}^3 \right\}. \label{action} \eea
It is invariant, in
the considered order, under the transformations \p{4transf}. It
differs from the sixth order of the $N=2$ BI action of \cite{Ket} just by
the last term. The same correction term was found in \cite{SKT} by
requiring self-duality and invariance under the bosonic $c, \bar c$
symmetry. We uniquely recovered it from the hidden $N=2$
supersymmetry. It would be interesting to inquire whether the 
equivalence of
these two sets of requirements persists to higher orders. Using the above
procedure, we in principle can restore the invariant action 
{\it to any order}.

\setcounter{equation}0
\section{N=4 Born-Infeld theory}
Finally, we derive the superfield equations of $N=4$ BI
theory within the same approach.

The $N=4, D=4$ Maxwell theory \cite{n4ym} is described
by the covariant strength superfield
${\cal W}_{ij}=-{\cal W}_{ji}, (i,j=1,\ldots,4)$, 
satisfying the following independent
constraints \cite{Sohn}
\bea
&&\bar{\cal  W}^{ij} \equiv \left( {\cal W}_{ij}\right)^* =
\frac{1}{2}\varepsilon^{ijkl}{\cal W}_{kl}\;,\label{realW} \\
&& D_{\alpha}^k {\cal W}_{ij}-\frac{1}{3}
\left(\delta_i^k D_{\alpha}^m {\cal W}_{mj} -
   \delta_j^k D_{\alpha}^m {\cal W}_{mi} \right) =0 \;.
\label{n8eomflat}
\eea
In contrast to the $N=2$ gauge theory, no off-shell
superfield formulation exists in the $N=4$ case: the constraints
\p{realW}, \p{n8eomflat} put the theory on shell.

As in the $N=2$ case, in order to construct a nonlinear generalization 
of \p{realW}, \p{n8eomflat}
one should firstly define the appropriate algebraic framework. It is given
by the following central charge-extended $N=8, D=4$ Poincar\'e superalgebra:
\bea
&& \left\{ Q_{\alpha}^i, {\bar Q}_{\dot{\alpha}j}
        \right\}=2\delta^i_jP_{\alpha\dot{\alpha}} \;, \quad
\left\{ S_{\alpha}^i, {\bar S}_{\dot{\alpha}j}
        \right\}=2\delta^i_jP_{\alpha\dot{\alpha}}\;, \nn
&& \left\{ Q_{\alpha}^i,  S_{\beta}^j
   \right\}=\varepsilon_{\alpha\beta} Z^{ij} \;, \quad
      \left\{ {\bar Q}_{\dot{\alpha}i}, {\bar S}_{\dot{\beta}j}
        \right\}=
 \varepsilon_{\dot{\alpha}\dot{\beta}}{\bar Z}_{ij} \;,
 \label{n8sa} \\
&& {\bar Z}_{ij}=\left(Z^{ij}\right)^* = 
\frac{1}{2}\varepsilon_{ijkl}Z^{kl}\;.\label{reality}
\eea
This is a $D=4$ notation for the type IIB Poincar\'e superalgebra 
in $D=10$.

We wish the $N=4, D=4$ supersymmetry $\left\{ P_{\alpha\dot{\alpha}},
 Q_{\alpha}^i, {\bar Q}_{\dot{\alpha}j}\right\}$ to remain unbroken, 
so we are
led to introduce the Goldstone superfields
\be\label{gfn8}
Z^{ij}\Rightarrow W_{ij}(x,\theta,\bar\theta)\;, \quad
S_{\alpha}^i\Rightarrow \psi^{\alpha}_i (x,\theta,\bar\theta)\;,\quad
{\bar S}_{\dot{\alpha}j}\Rightarrow 
{\bar\psi}_i^{\dot\alpha}(x,\theta,\bar\theta)\;.
\ee
The reality property \p{reality} automatically implies the
constraint \p{realW} for $W_{ij}$:
\be
\bar W^{ij} = \frac{1}{2}\varepsilon^{ijkl} W_{kl}~. \lb{real1}
\ee

On the coset element $g$
\be\label{n8coset}
g= \mbox{exp}\,i\left(-x^{\alpha\dot{\alpha}}P_{\alpha\dot{\alpha}}
   +\theta^{\alpha}_i Q^i_{\alpha}+{\bar\theta}^i_{\dot\alpha}
    {\bar Q}_i^{\dot\alpha}\right)
  \mbox{exp}\,i\left(\psi^{\alpha}_i S^i_{\alpha}+{\bar\psi}^i_{\dot\alpha}
    {\bar S}_i^{\dot\alpha} + W_{ij}Z^{ij} \right)
\ee
one can realize the entire $N=8, D=4$ supersymmetry \p{n8sa} by left
shifts. The Cartan forms (except for the central charge one) and covariant
derivatives formally coincide with \p{vectorcf}-\p{algebracovd}, the indices
$\left\{ i,j\right\}$ now ranging from 1 to 4. The central charge Cartan
form reads:
\be\label{n8ccform} \omega^Z_{ij}=dW_{ij}+\frac{1}{2}\left(
d\theta^{\alpha}_i\psi_{\alpha j}-
 d\theta^{\alpha}_j\psi_{\alpha i}+
\varepsilon_{ijkl}d{\bar\theta}^k_{\dot\alpha}
    {\bar\psi}{}^{\dot\alpha l}\right)~. \;
\ee
By construction, it is covariant under all transformations of the
$N=8, D=4$  Poincar\'{e} supergroup. The Goldstone superfields 
$\psi_{\alpha i}$ and
${\bar\psi}^k_{\dot\alpha}$ can be covariantly eliminated by the 
inverse Higgs procedure,
as in the previous case. The proper constraint reads as follows:
\be\label{n8con1}
\omega^Z_{ij} \vert_{d\theta, d\bar\theta} =0 \;.
\ee
It amounts to the following set of equations:
\be
(a)\;\cD_{\alpha}^k W_{ij}+\frac{i}{2}\left( \delta_i^k \psi_{\alpha j}-
     \delta_j^k\psi_{\alpha i} \right) =0\;, \quad
(b) \; \cDb_k^{\dot\alpha}W_{ij}+\frac{i}{2}
\varepsilon_{ijkl}{\bar \psi}^{\dot\alpha l}
   =0 \;,  \label{eq2}
\ee
which are actually conjugated to each other in virtue of \p{real1}. We
observe that, besides expressing the fermionic Goldstone superfields through
the basic bosonic one $W_{ij}$:
\be\label{eq3} \psi_{\alpha
i}=-\frac{2i}{3}\cD_{\alpha}^jW_{ij}\;, \quad {\bar \psi}^{\dot\alpha
i}=-\frac{2i}{3}\cDb_j^{\dot\alpha}{\bar W}^{ij} \;,
\ee
eqs. \p{eq2} impose the nonlinear constraint
\be\label{n8eom}
\cD_{\alpha}^k W_{ij}-\frac{1}{3}\left(\delta_i^k \cD_{\alpha}^m W_{mj} -
   \delta_j^k\cD_{\alpha}^m W_{mi} \right) =0 
\ee
(and its conjugate). This is the sought nonlinear generalization 
of \p{n8eomflat}.

It is straightforward to show that eq. \p{n8eom} implies the disguised form
of the BI equation \p{case1eom} for the nonlinear analog of the abelian 
gauge field strength. For the six physical bosonic fields $W_{ij}\vert$
we expect the equations corresponding to the static-gauge of
$3$-brane in $D=10$ to hold. Thus eqs. \p{real1}, \p{n8eom} plausibly 
give a manifestly worldvolume
supersymmetric description of D3-brane in a flat $D=10$ Minkowski background.
No simple off-shell action can be constructed in this case, since 
such an action
is unknown even for the free $N=4$ Maxwell theory. But even the construction
of the physical fields component action for this $N=4$ BI theory is of
considerable interest. We hope to study this system in more detail
elsewhere.

\section{Conclusions}
In this paper, generalizing the nonlinear-realizations approach of
\cite{BIK3}, we constructed superfield equations describing $N=2$ and $N=4$
supersymmetric BI theories with extra nonlinearly realized $N=2$ and $N=4$
supersymmetries. These systems, by construction, realize a $1/2$ partial
breaking of the appropriate central-charge extended $N=4$ and $N=8$, $D=4$
Poincar\'e supersymmetries which are in fact a $D=4$ form of $N=2$
supersymmetries in $D=6$ and $D=10$. Thus, the equations constructed
admit a natural interpretation as providing a manifestly supersymmetric
worldvolume description of D3-branes in $D=6, 10$. In both cases
the basic objects are the bosonic Goldstone superfields associated with the
central-charge generators. They are nonlinear analogs of the $N=2$ and $N=4$
Maxwell superfield strengths.

Besides tasks for a future study, such as the construction of the full
invariant actions for the considered systems (an off-shell action for $N=2$
BI and an on-shell one for $N=4$ BI) and deducing super BI theories
in higher dimensions (e.g., $N=(1,0)$, $D=6$ BI
theory associated with a nonlinear realization of $N=(2,0)$, $D=6$
supersymmetry \cite{BIK1,BIK2}), let us mention a more ambitious problem.
The above consideration raises the natural question as to how to
make the nonlinear-realizations approach suitable also for deriving
the equations of supersymmetric {\it non-abelian} BI theories with hidden
supersymmetries (see \cite{Bergsh} for a discussion of
such systems in the Green-Schwarz
formulation). Since the gauge superfield strengths always
appear as Goldstone superfields in the nonlinear-realizations approach
(associated with the spinor generators in the $N=1, D=4$ case \cite{BG2}
and  the central-charge ones in the $N=2$ and $N=4$ cases), 
in order to be able
to treat a non-abelian covariant superfield strength
in a similar way, it seems necessary to pass to a new kind
of superalgebras with the generators taking values
in the gauge group algebra. Thus these generalized superalgebras should
provide a non-trivial unification of supersymmetry with the gauge groups.
The non-abelian analogs of the equations constructed here
must inevitably involve the gauge connections for which there should also
be a natural geometric place in nonlinear realizations of the hypothetical
generalized supersymmetries. The treatment of gauge fields as the
Goldstone fields \cite{IO} could be suggestive in this respect.

\vspace{0.3cm}
\noindent{\bf Acknowledgements.}\hskip 1em
We are grateful to Eric Bergshoeff,
Sergio Ferrara, Sergei Ketov, Sergei Kuzenko,
Dmitri Sorokin and Mario Tonin for useful discussions at different stages of
this work. E.A. and S.K. thank LNF INFN for warm hospitality. This work was
supported in part by the Fondo Affari Internazionali Convenzione Particellare
INFN-JINR, grants RFBR-CNRS 98-02-22034, RFBR-DFG-99-02-04022, 
RFBR 99-02-18417 and NATO Grant PST.CLG 974874.

\end{document}